\documentclass[aoas,preprint]{imsart}

\RequirePackage{amsthm,amsmath,amsfonts,amssymb}
\RequirePackage[authoryear]{natbib}

\startlocaldefs
\usepackage{amsmath}
\usepackage{graphicx}
\usepackage{enumerate}
\usepackage{url} 
\usepackage{adjustbox}
\usepackage{bigints}
\usepackage[ruled, vlined, linesnumbered]{algorithm2e}
\usepackage[american]{babel}
\usepackage{amsthm}
\usepackage{amssymb}
\usepackage{color}
\usepackage{float}
\usepackage{mathtools}
\usepackage{subfig}

\providecommand{\tabularnewline}{\\}
\floatstyle{ruled}
\newfloat{algorithm}{tbp}{loa}
\providecommand{\algorithmname}{Algorithm}
\floatname{algorithm}{\protect\algorithmname}

\numberwithin{equation}{section}
\theoremstyle{plain}
\newtheorem{lem}{\protect\lemmaname}
\theoremstyle{plain}
\newtheorem{thm}{\protect\theoremname}
\theoremstyle{remark}
\newtheorem*{rem*}{\protect\remarkname}
\theoremstyle{plain}

\newtheorem{prop}{Proposition}

\newtheorem*{assumption*}{\assumptionnumber}
\providecommand{\assumptionnumber}{}
\makeatletter
\newenvironment{assumption}[2]
 {%
  \renewcommand{\assumptionnumber}{\bf{Assumption #1}}%
  \begin{assumption*}%
  \protected@edef\@currentlabel{#1}%
 }
 {%
  \end{assumption*}
 }
\makeatother

\newtheorem*{theorem*}{\theoremnumber}
\providecommand{\theoremnumber}{}


\SetKwInput{KwInput}{Input}
\SetKwInput{KwOutput}{Output}
\let\oldnl\nl
\newcommand{\nonl}{\renewcommand{\nl}{\let\nl\oldnl}}

\DeclarePairedDelimiter{\nint}\lceil\rceil

\DeclareMathOperator*{\argmin}{arg\,min}

\providecommand{\lemmaname}{Lemma}
\providecommand{\remarkname}{Remark}
\providecommand{\theoremname}{Theorem}

\endlocaldefs

\begin{document}

\begin{frontmatter}
\title{Robust Causal Inference for Incremental Return on Ad Spend with
Randomized Paired Geo Experiments}
\runtitle{Robust Causal Inference for iROAS}

\begin{aug}
\author[]{\fnms{Aiyou} \snm{Chen}}
\and
\author[]{\fnms{Timothy C.} \snm{Au}}
\address[]{aiyouchen@google.com and timau@google.com\\
  Google LLC}

\end{aug}

\begin{abstract}
  Evaluating the incremental return on ad spend (iROAS) of a prospective online
marketing strategy (i.e., the ratio of the strategy's causal effect on some
response metric of interest relative to its causal effect on the ad spend) has
become increasingly more important.
Although randomized ``geo experiments'' are frequently employed for this
evaluation, obtaining reliable
estimates of iROAS can be challenging as oftentimes only a small number of
highly heterogeneous units are used.  Moreover, advertisers frequently impose
budget constraints on their ad spends, which further complicates
causal inference by introducing interference between the experimental units.
In this paper, we formulate a novel statistical framework for inferring the
iROAS of online advertising from randomized paired geo experiment which
further motivates and provides new insights into Rosenbaum's arguments on
instrumental variables, and we propose and develop a robust,
distribution-free and interpretable estimator ``Trimmed Match'', as well as a
data-driven choice of the tuning parameter which may be of
independent interest.
We investigate the sensitivity of Trimmed Match to some violations of its
assumptions and show that it can be more efficient than some alternative
estimators based on simulated data. We then demonstrate its practical utility
with real case studies.

\end{abstract}


\end{frontmatter}

\section{Introduction}
\label{sec:intro}

Similar to traditional offline media such as television, radio and print, the
primary goal of online advertising is to help promote the selling of goods and
services.  However, despite these shared goals, online advertising has been the
leading source of advertising revenue in the United States since 2016
\citep{iab:2018}.  \citet{goldfarb:tucker:2011}
attribute this success of online advertising to its superiority over
other media in terms of its measurability and targetability.

A prospective online marketing strategy (e.g., expanding the list of search
keywords on which to advertise) is frequently evaluated
in terms of its incremental return on ad spend (iROAS)---that is, the ratio of
the strategy's causal effect on some response metric of interest relative to its
causal effect on the ad spend. Here the response metric of interest may be
for example, revenue from online sales, offline sales,
or overall sales which may be affected by
the ad. Indeed, this evaluation
has become progressively more important as advertisers increasingly seek to
optimize the impact of their marketing decisions---an evaluation that is, in
theory, facilitated by large-scale randomized
experiments (i.e. ``A/B tests'') which randomize users
to different ad serving conditions
\citep{goldfarb:tucker:2011,johnson2017ghost}.
In practice, however, privacy concerns which restrict the collection of
user data and technical issues such as cookie churn
and multiple device usage have made it hard
to maintain the integrity of a randomized user experiment in the online
advertising context 
\citep{gordon:etal:2018}.
Consequently, observational studies remain an area of active
research for estimating the causal impact of online marketing strategies (e.g.,
\citet{varian2016causal}, \citet{sapp2017near}, \citet{chen2018bias},
and the references therein), although the empirical studies of
\citet{lewis:rao:2015} and \citet{gordon:etal:2018} continue to suggest caution
when using observational methods despite these recent advances.

Indeed, randomized experiments are still regarded as the ``gold standard''
for causal inference \citep{imbens:rubin:2015} and, to mitigate
some of the challenges of user-level experimentation, advertisers frequently
instead employ randomized ``geo experiment'' designs \citep{vaver:koehler:2011}
which partition a geographic region of interest into a set of
non-overlapping ``geos'' (e.g., Nielsen Media Research's 210 Designated
Market Areas\footnote{\url{https://www.nielsen.com/intl-campaigns/us/dma-maps.html}}
which subdivide the United States) that are regarded as the units of experimentation
rather than the individual users themselves.
More formally, let
$\mathcal{G}$ be the set of geos in a target population where, for a geo
$g\in\mathcal{G}$, we let $(S_{g},R_{g})\in\mathcal{R}^{2}$ denote
its observed bivariate outcome with ad spend $S_{g}$ and
response $R_{g}$.
Following the Neyman-Rubin causal framework,
we denote geo $g$'s potential
outcomes under the control and treatment ad serving conditions as
\begin{align*}
(S_{g}^{(C)},R_{g}^{(C)}) & \text{ and }(S_{g}^{(T)},R_{g}^{(T)}),
\end{align*}
respectively, where we can only observe one of these two bivariate potential
outcomes for each geo $g$.  Therefore, relative to the control condition, there
are two unit-level causal effects caused by the treatment condition
for each geo $g$---the incremental ad spend and the incremental response which
are defined by
\begin{align*}
S_{g}^{(T)}-S_{g}^{(C)} & \text{ and }R_{g}^{(T)}-R_{g}^{(C)},
\end{align*}
respectively.
However, advertisers frequently
find the iROAS, i.e. the ratio of incremental response to incremental ad spend, to be a more informative and actionable measure of advertising performance:
\begin{align}
\theta_{g} & =\frac{R_{g}^{(T)}-R_{g}^{(C)}}{S_{g}^{(T)}-S_{g}^{(C)}},\label{eq:theta-geo}
\end{align}
for $g \in \mathcal{G}$.
Following \citet{kerman:etal:2017} and \citet{kalyanam:etal:2018}, and
letting $\left|\cdot\right|$ denote the set cardinality function,
the overall iROAS with respect to $\mathcal{G}$ can be
defined as the ratio of the average incremental response to the average
incremental ad spend:
\begin{align}
\theta^{*} & =\frac{\frac{1}{|\mathcal{G}|}\sum_{g\in\mathcal{G}}{\left(R_{g}^{(T)}-R_{g}^{(C)}\right)}}{\frac{1}{|\mathcal{G}|}\sum_{g\in\mathcal{G}}{\left(S_{g}^{(T)}-S_{g}^{(C)}\right)}},\label{eq:theta}
\end{align}
which is the parameter of primary interest in this paper.

In a randomized experiment,
one can use the group difference 
to obtain unbiased estimates
of the average incremental response and average incremental ad spend.  The ratio
of these group differences 
then gives an empirical estimate of $\theta^{*}$:
\begin{align}
\hat{\theta}^{(emp)} & =\frac{\frac{1}{|\mathcal{T}|}\sum_{g\in\mathcal{T}}R_{g}-\frac{1}{|\mathcal{C}|}\sum_{g\in\mathcal{C}}R_{g}}{\frac{1}{|\mathcal{T}|}\sum_{g\in\mathcal{T}}S_{g}-\frac{1}{|\mathcal{C}|}\sum_{g\in\mathcal{C}}S_{g}}\label{eq:random-empirical}
\end{align}
where $\mathcal{T}$ and $\mathcal{C}$ denote the set of geos in treatment and in
control, respectively.
Similar empirical estimates have been commonly used for
effect ratios in other applications, e.g., the incremental cost-effectiveness
ratio which summarizes the cost-effectiveness of a health care intervention
\citep{chaudhary.stearns.1996,bang.zhao.2014}.

However, geo experiments often introduce some additional complexity which makes
the causal estimation of the iROAS more difficult. The complexity can be
attributed mostly to two sources of interference:
1) spillover effects (e.g., from consumers traveling across geo boundaries), and
2) budget constraints on ad spend.
Existence of interference, if not handled properly, may invalidate traditional
causal inference which relies on the ``stable unit treatment value assumption''
(SUTVA)---that is, the presumption that the treatment applied to one unit does
not affect the outcome of another unit \citep{rubin:1980}.

Interference due to the first source can be controllable to a large extent by
using bigger geographically clustered regions as the experimental units
\citep{vaver:koehler:2011,rolnick.etc.2019} and, therefore, such interference is assumed to be
ignorable in this paper.
As the consequence, however, geo experiments frequently involve only a small
number of geos \citep{vaver:koehler:2011} where the
distributions of $\{S_{g}:g\in\mathcal{G}\}$ and $\{R_{g}:g\in\mathcal{G}\}$
can be very heavy-tailed and, as a result, the empirical estimator defined
in equation \eqref{eq:random-empirical} can be very unreliable.

Interference due to the second source is more subtle and less controllable.
Although advertisers have
some control over the online marketing strategies that they employ
(e.g., which search keywords to advertise on, the maximal price they are willing
to pay to show their ads, etc.), they are competing in a dynamic
ecosystem where the actual delivery of their online
ads is determined by advertising platforms which run auctions and use machine
learning models to optimize the ad targeting in real-time to maximize key
performance indicators such as clicks, site visits, and purchases
\citep{varian:auction:2009,johnson2017ghost}. Thus, for a given budget
constraint for geos in the treatment group (or the control group),
these advertising platforms may choose to
allocate ad spend to one geo at the expense of others which, in turn,
introduces interference in the responses observed in each geo.
We discuss later that this interference can be handled naturally for various
kinds of geo experiments, unlike interference which has recently been studied
elsewhere (e.g., medical science, social network), see
\citet{rosenbaum:2007:interference}, \citet{luo:etc:2012:interference},
\citet{athey:eckles:imbens:2018:intereference} and references therein.

The key contributions of this paper are: 1) the formulation of a novel
statistical framework for inferring the iROAS $\theta^{*}$ from randomized
paired geo experiments which embed the complex nature of small sample sizes,
data heavy-tailedness, and interference due to budgetary constraints,
2) the proposal and development of a robust and distribution-free
estimator ``Trimmed Match'' with a simple interpretation, and
3) extensive simulations and real case studies demonstrating that Trimmed Match
can be more efficient than some alternative estimators.

The rest of this paper is organized as follows. We first provide the background
on geo experiments in the online advertising context in Section
\ref{sec:Background}.  Afterwards, we formulate a statistical framework for
inferring $\theta^{*}$ in a randomized paired geo experiment in Section
\ref{sec:A-model-free-framework}. Under this framework and in the spirit of
\citet{rosenbaum1996identification, rosenbaum2002covariance}, we review a few
distribution-free estimators based on common test statistics in Section
\ref{sec:binomial-test}, and we propose the Trimmed Match estimator in Section
\ref{sec:trimmed-match}.
Section \ref{sec:choice-trim-rate} introduces a data-driven choice of the trim
rate. Simulations demonstrating the robustness and
efficiency of Trimmed Match are presented in Section
\ref{sec:Simulation-and-sensitivity}, and some real case studies illustrating
the performance of Trimmed Match in practice are shown in Section
\ref{sec:case-studies}.
Finally,
Section \ref{sec:Discussion} concludes with some suggestions for future
research.
Fast computation of Trimmed Match is presented in Appendix \ref{sec:trimmed-match-comp}.

Trimmed Match has been applied for advertiser studies at Google. The Python
library for the implementation is available at GitHub \citep{tmcode}.

\section{Background\label{sec:Background}}

Geo experiments are now a standard tool for the causal
measurement of online advertising at Google---see, for example, \citet{blake:etal:2015},
\citet{ye:etal:2016}, and \citet{kalyanam:etal:2018}.\footnote{For a list of
``geo targets'' supported by Google AdWords, see:
\\
 \indent\indent\indent \url{https://developers.google.com/adwords/api/docs/appendix/geotargeting}}
There has been some related work in terms of estimating the effectiveness of
online advertising with geo experiments, but to the best of our knowledge,
all work to date has been model-based.

Notably, after introducing the concept of a randomized geo experiment design
for online advertising, \citet{vaver:koehler:2011} proceed to analyze them
using a two stage weighted linear regression approach, called Geo-Based
Regression (GBR).
The first stage fits a linear predictive model for the geo-level potential control
ad spend using data from the control group, where pre-experimental ad spend is
used as regressors and model weights to
try to account for heteroscedasticity caused by the differences in geo size.
The second stage of GBR fits a regression for the response
variable, where pre-experimental response and incremental ad spend---which
is 0 by definition for geos in the control group, but for geos in the treatment
group is inferred by the difference between the observed ad spend and the counterfactual
ad spend predicted from the first stage---are used as regressors and model weights to
try to account for heteroscedasticity, and the iROAS parameter is the
coefficient of incremental ad spend.

More recently, to address situations where there are only a few geos available
for experimentation, \citet{kerman:etal:2017}
propose a Time-Based Regression (TBR) approach which uses a constrained version
of the Bayesian structural time series model
\citet{brodersen:etal:2015}
to estimate the overall incremental response for the treatment group,
where the control group's time series is used to contemporaneously model the
treatment group's ``business as usual'' behavior prior to the experiment, and
then subsequently used in conjunction with the trained model to predict what
the treatment group's ``business as usual'' counterfactual would have been had
the experiment not occurred.

However, it can be shown that these methods rely on some strong modeling
assumptions that are often hard to justify in practice. For GBR, the result can be
quite sensitive to the choice of weights, and furthermore,
even if the geo-level incremental ad spends are known,
unlike more recent regression adjustment models
\citep{lin2013agnostic}, its second stage regression
may still suffer from the endogeneity problem---incremental ad spend
may correlate with the residual---despite randomness in the treatment
assignment. As a natural extension of GBR, one might attempt to
fit the heterogeneous geo-level causal effects on the response and ad spend
separately using parametric or nonparametric models
\citep{bloniarz2016lasso,WagerandAthey2018,kunzel2019metalearners}. Besides the
requirement of larger sample size, however, this
may not be straight-forward since budget constraints on ad spend may
break the assumption of independent measurements behind the models.
Meanwhile, TBR assumes a stable linear relationship regarding the
contemporaneous ``business as usual'' time series
between the treatment group and the control group from the
pre-experimental period into the experimental period. But this is an untestable
assumption and
may not hold in practice. Given the temporal dynamics (e.g., the COVID-19 pandemic
as an extreme case),
it is important to have a method which is robust and does not rely on any
fragile and untestable modeling assumptions.
Such a method is especially desirable when it needs to be built into
a product to serve many geo experiments seamlessly.

\section{A Statistical Framework for Inferring the iROAS}
\label{sec:A-model-free-framework}

We first consider the scenario where there is no budgetary constraint so that
there is no interference between geos.
Recall from \eqref{eq:theta-geo} that a geo $g$'s unit-level
iROAS $\theta_{g}$ is defined in terms of the ratio of its incremental
response to its incremental ad spend. Rearranging the
terms in this definition then leads to the following lemma, which
serves as the basis for our statistical framework.
\begin{lem}
$R_{g}^{(T)}-\theta_{g}S_{g}^{(T)}=R_{g}^{(C)}-\theta_{g}S_{g}^{(C)}$
for every geo $g\in\mathcal{G}$. \label{lem:important-observation}
\end{lem}
Lemma \ref{lem:important-observation} implies that the quantity
\begin{align*}
Z_g &\equiv R_{g}-\theta_{g}S_{g}, 
\end{align*}
which is not observable due to unknown $\theta_g$,
remains the same regardless of whether geo $g$ is assigned to treatment
or control. Loosely speaking, the quantity $Z_g$ measures geo $g$'s ``uninfluenced
response''---that is, the part of $g$'s baseline response due to, for example,
seasonality in the market demand which is not influenced by its ad spend.

As previously discussed in Section \ref{sec:intro}, budgetary constraints
may introduce complex interference and thus a violation of SUTVA
since the ad spend allocated to one geo
may come at the expense of others; therefore the experimental outcome
$(S_g, R_g)$ for each geo $g$ may be affected by the treatment assignment of
other geos. In particular, for a design on $n$ matched pairs, there are $2^n$ possible
assignments each associated with its own potential outcome vector of length $2n$,
where the realized potential outcome vector depends on the materialized assignment---
see for instance \cite{hudgens2008toward} for a detailed formulation.
But in light of Lemma \ref{lem:important-observation}, in this paper we assume
for notational simplicity a relaxed version of SUTVA
when there is interference due to budgetary constraints,
which is formally described as Assumption \ref{assm:SUTVA} below.

\begin{assumption}{0}{}\label{assm:SUTVA}
The uninfluenced response $Z_g$ for any geo $g$ introduced by
Lemma \ref{lem:important-observation} is invariant to
both its own treatment assignment and the treatment assignment of other geos.
\end{assumption}

It is not hard to verify that under Assumption \ref{assm:SUTVA}, the parameters
$\theta_g$ in \eqref{eq:theta-geo}
are well defined.
Assumption \ref{assm:SUTVA} trivially holds when there is no interference
(e.g., if the advertiser had no budget constraints or if the total actual ad
spend was below the pre-specified budget constraint).  More generally,
by the decomposition
\begin{align*}
R_g &= (R_g - \theta_g S_g) + \theta_g S_g \\
    & = Z_g + \theta_g S_g,
\end{align*}
it is important to note that under Assumption \ref{assm:SUTVA}, $Z_g$ is not affected
by geo assignment and thus any interference introduced from budgetary
constraints only affects the response through the magnitude of materialized
ad spend $S_g$ and iROAS $\theta_g$. In other words, this specifies a simple linear form which quantifies how the
interference due to budgetary constraints affects the measurements. Consequently,
Assumption \ref{assm:SUTVA} may still hold
in less trivial situations with budget constraints such as:
\begin{itemize}
\item When the potential ad spend under the control condition is known to be 0
for every geo (e.g., showing online ads under the treatment condition versus not
showing ads under the control condition), or when the potential ad spend under
the treatment condition is known to be 0 for every geo (e.g., ``go dark`` experiments
where no ads are shown under the treatment condition \citep{blair2004better}).
\item When advertisers use their budget constraint to pre-specify each geo's
potential ad spend under the control condition (but not necessarily when under
the treatment condition), or when advertisers pre-specify each geo's potential
ad spend under the treatment condition (but not necessarily when under the
control condition).
\end{itemize}
These scenarios encompass many of the geo experiment designs in practice and,
consequently, we assume that Assumption \ref{assm:SUTVA} holds in the remainder
of this paper.

\begin{prop}
\label{thm:random-experiment}In a completely randomized geo experiment,
under Assumption \ref{assm:SUTVA},
the distribution of $R_{g}-\theta_{g}S_{g}$ is the same between the
treatment group and the control group.
\end{prop}
Proposition \ref{thm:random-experiment}, whose proof directly follows from
Assumption \ref{assm:SUTVA}, provides
a general framework for inferring the unit-level iROAS $\{\theta_g:g\in \mathcal{G}\}$
(e.g., by parameterizing $\theta_g$ with geo-level features)
and the overall target population iROAS $\theta^*$ by simplifying
the bivariate causal inference problem to a single dimension.
In this paper, we formulate a statistical framework specifically for inferring
$\theta^{*}$ by assuming, just as \citet{vaver:koehler:2011} do, that the unit-level
iROAS $\theta_{g}$ are all identical.

\begin{assumption}{1}{}\label{assm:common_iroas}
$\theta_{g}=\theta^{*}$ for all geos $g\in\mathcal{G}$.
\end{assumption}

Although the rigorous verification of Assumption \ref{assm:common_iroas}
is beyond the scope of this paper, our sensitivity analysis in Section
\ref{sec:Simulation-and-sensitivity} suggests that estimates of $\theta^{*}$ can
still be reliable even if the unit-level iROAS $\theta_{g}$ moderately differ, while
our hypothesis tests in Section \ref{sec:case-studies}
indicate that this assumption is compatible with data observed from
several real case studies.

Following the recommendations of \citet{vaver:koehler:2011}, in the
remainder of this paper we consider a randomized paired design where $2n$
geos are matched into $n$ pairs prior to the experiment such that, within
each pair, one geo is randomly selected for treatment and the other geo for
control.

Let $A_i$ be the random assignment for the two geos in the
$i$th pair
with $P(A_i=1)=P(A_i=-1)=\frac{1}{2}$, where $A_i=1$ indicates that the 1st geo
receives treatment and the 2nd geo receives control, while $A_i=-1$ indicates that the 2nd
geo receives treatment and the 1st geo receives control.
Let $(S_{i_1}, R_{i_1})$ and $(S_{i_2}, R_{i_2})$ be the observed spend and response values
for the 1st geo and the 2nd geo respectively in the $i$th pair.
Let $X_i$ and $Y_i$ be the observed differences in the ad spends and responses, respectively,
between the treatment geo and the control geo in the $i$th pair, that is,
\begin{align}
X_{i}=(S_{i_1}-S_{i_2})\cdot A_i & \text{ and } Y_{i}=(R_{i_1}-R_{i_2})\cdot A_i. \label{eq:paired-diff}
\end{align}
Let
\begin{align}
\epsilon_{i}(\theta) & =Y_{i}-\theta X_{i}.\label{eq:epsilon}
\end{align}
Table \ref{tab:pair} lists the notation and definitions used for the
$i$th pair of geos.

\begin{table}[t]
\centering{}\caption{\label{tab:pair}Description of the notation used for the $i$th pair
of geos.}
\begin{tabular}{c|l}
\hline
Notation  & Description \tabularnewline
\hline
$S_{i_1}$, $R_{i_1}$  & Observed ad spend and response for the 1st geo \tabularnewline
$S_{i_2}$, $R_{i_2}$  & Observed ad spend and response for the 2nd geo \tabularnewline
$A_i$                 & Indicator which geo receives treatment or control \tabularnewline
$X_{i}=(S_{i_1}-S_{i_2}) \cdot A_i $  & Spend difference between treatment and control \tabularnewline
$Y_{i}=(R_{i_1}-R_{i_2}) \cdot A_i$  & Response difference between treatment and control \tabularnewline
$\epsilon_{i}(\theta)=Y_{i}-\theta X_{i}$  & Difference in the ``uninfluenced responses'' with respect to $\theta$ \tabularnewline
\hline
\end{tabular}
\end{table}

\begin{prop}
\label{thm:symmetry}With a randomized paired design for geo experiments,
under Assumption \ref{assm:SUTVA} \& \ref{assm:common_iroas}, $\{\epsilon_i(\theta^*): i=1,\ldots, n\}$
are mutually independent and
the distribution of $\epsilon_{i}(\theta^{*})$
is symmetric about 0 for $i = 1,\ldots,n$.
\end{prop}

\begin{proof}
Let $Z_{i_1}$ and $Z_{i_2}$ be the uninfluenced responses associated with the two
geos in the $i$th pair, i.e. $Z_{i_j} = R_{i_j} - \theta^* S_{i_j}$ for $j=1,2$.
Then we have
\begin{align*}
\epsilon_{i}(\theta^{*}) & = (R_{i_1}-R_{i_2}) \cdot A_i - \theta^* \cdot (S_{i_1} - S_{i_2}) \cdot A_i \\
                         & = (Z_{i_1} - Z_{i_2})\cdot A_i.
\end{align*}
Under Assumption \ref{assm:SUTVA}, $Z_{i_1}$ and $Z_{i_2}$ are non-random quantities and
are invariant to any treatment assignment of the $2n$ geos. The conclusion
follows immediately since $\{A_i: i=1, \ldots, n\}$ are i.i.d. and each $A_i$ is symmetric about 0.
\end{proof}

Proposition \ref{thm:symmetry} provides a general framework that facilitates
the estimation of $\theta^{*}$---regardless of how complicated
the bivariate distribution of $\{(R_{g},S_{g}):g\in\mathcal{G}\}$
may be, we can always reformulate the causal inference problem in terms
of a simpler univariate ``location'' problem that is defined in terms of the
symmetric distribution of each $\epsilon_{i}(\theta^{*})$.

By Proposition \ref{thm:symmetry}, the average of $\{\epsilon_{i}(\theta^{*}):1\leq i\leq n\}$
is expected to be 0, so by setting
\begin{align*}
\frac{1}{n}\sum_{i=1}^{n}\epsilon_{i}(\theta) & =0
\end{align*}
and then solving for $\theta$, we arrive at the following estimator for
$\theta^{*}$:
\begin{align}
\hat{\theta}^{(emp)} & =\frac{\sum_{i=1}^{n}Y_{i}}{\sum_{i=1}^{n}X_{i}},\label{eq:empirical}
\end{align}
which coincides with, and also further motivates, the empirical estimator
given in \eqref{eq:random-empirical} with $|\mathcal{T}|=|\mathcal{C}|$.
However, recall from our discussions in Section \ref{sec:intro} that the empirical
estimator may be unreliable when the bivariate distribution of $\left\{ \left(R_{g},S_{g}\right):g\in\mathcal{G}\right\} $
is heavy tailed.

Although the iROAS estimation problem is fundamentally different from the
classical location problem as studied extensively in the statistics literature,
the reformulation of the problem in terms of the symmetry of the
$\epsilon_{i}(\theta^{*})$ values about 0 facilitates the application of robust
statistical methods to address the heterogeneity issue of geo experiments.
For conciseness, we only consider three such techniques in this paper
and leave the exploration of other robust statistical methods to future
work---we refer the reader to \citet{tukey1963less},
\citet{lehmann1975nonparametrics}, and \citet{huber2009robust} for a
comprehensive overview of such techniques.  Specifically, we first briefly
review the application of binomial sign test and the Wilcoxon signed-rank test
in Section \ref{sec:binomial-test}. Afterwards, in Section
\ref{sec:trimmed-match}, we develop a robust and more easily interpretable
estimator based on the trimmed mean, and we demonstrate its efficiency in
practice through simulations and real case studies presented in Sections
\ref{sec:Simulation-and-sensitivity} and \ref{sec:case-studies}.

\section{Related work \label{sec:binomial-test}}

A similar statistic in the form of \eqref{eq:epsilon} was first proposed and
studied by
\citet{rosenbaum1996identification, rosenbaum2002covariance}
to generalize an instrumental variable argument made by
\citet{angrist1996identification}, but in a different context and without
the concern of interference as studied in this paper. In this section, we
review two distribution and covariate free estimators of $\theta^{*}$ along the
same lines and refer to \citet{rosenbaum.book.2019} for a more comprehensive overview.

For any $\theta\in\mathcal{R},$ let $M_n(\theta)$ be a statistic for testing
symmetry where, in the case of binomial sign test, we have
\begin{align*}
M_{n}(\theta) & =\sum_{i=1}^{n}\left(I(\epsilon_{i}(\theta)>0) - \frac{1}{2}\right),
\end{align*}
with $\epsilon_{i}(\theta)$ given by \eqref{eq:epsilon} and where $I(\cdot)$ is
the indicator function, while in the case of Wilcoxon's signed-rank test
we have
\begin{align*}
  M_{n}(\theta) & = \sum_{i=1}^{n} \text{sgn}(\epsilon_i(\theta)) \cdot \text{rank}(|\epsilon_i(\theta)|),
\end{align*}
with $\text{sgn}(\cdot)$ and $\text{rank}(\cdot)$ denoting the sign and rank
functions, respectively.  We refer the reader to
\cite{lehmann1975nonparametrics} for additional details on tests of
symmetry.

\begin{prop}
  \label{thm:binomial}Under Assumption \ref{assm:SUTVA} \& \ref{assm:common_iroas}, the test
  statistic $M_{n}(\theta^{*})$ for either binomial sign test or Wilcoxon's
  signed-rank test follow a known distribution that is symmetric about 0.
\end{prop}

Proposition \ref{thm:binomial}, whose proof follows directly from Proposition
\ref{thm:symmetry}, allows us to construct a $100(1 - \alpha)\%$ confidence
interval for $\theta^{*}$---if we let $q_{1-\alpha/2}$ be the $(1-\alpha/2)$
quantile for the distribution of $M_n(\theta^*)$ under Assumption
\ref{assm:common_iroas}, then we identify the minimal interval containing all
$\theta\in\mathcal{R}$ satisfying
$|M_{n}(\theta)|\leq q_{1-\alpha/2}.$

Note, however, that $M_n(\theta)=0$ may not always have a root. Following
\cite{rosenbaum1996identification}, the point estimator of $\theta^{*}$ can be
defined as the average of the smallest and largest values of $\theta$ that
minimize $|M_n(\theta)|$---that is,
\begin{align}
\hat{\theta} & =\frac{\inf\Theta_{M}+\sup\Theta_{M}}{2}, \label{eq:binomial}
\end{align}
where
$\Theta_{M} = \argmin_{\theta \in \mathcal{R}}|M_n(\theta)|$.
In the remainder of this paper, we let $\hat{\theta}^{(binom)}$ and
$\hat{\theta}^{(rank)}$ denote the estimators which correspond to binomial
sign test statistic and Wilcoxon's signed-rank test statistic, respectively.

\section{The ``Trimmed Match'' Estimator \label{sec:trimmed-match}}

In this section, we derive an important estimator for $\theta^{*}$
based on the trimmed mean under Proposition \ref{thm:symmetry}.
In particular, for a randomized paired geo experiment, let $\{(X_{i},Y_{i}):1\leq i\leq n\}$
be as defined in \eqref{eq:paired-diff} and, for any $\theta\in\mathcal{R}$, let
$\{\epsilon_{i}(\theta):1\leq i\leq n\}$ be as defined in \eqref{eq:epsilon} with
the corresponding order statistics given by
$\epsilon_{(1)}(\theta)\leq\epsilon_{(2)}(\theta)\leq\ldots\leq\epsilon_{(n)}(\theta)$.

\subsection{Point Estimation}
\label{subsec:trimmed-match-point}

For a fixed value $\lambda\in[0, 1/2)$, the trimmed mean
statistic as a function of $\theta$ is defined as:
\begin{align}
\overline{\epsilon}_{n\lambda}(\theta) & \equiv\frac{1}{n-2m}\sum_{i=m+1}^{n-m}\epsilon_{(i)}(\theta),\label{eq:Tmean}
\end{align}
where $m\equiv\nint{n\lambda}$ is the minimal integer greater or
equal to $n\lambda$. Here $\lambda$ is a tuning parameter which is commonly
referred to as the trim rate and, in order to be well defined, $\lambda$ must satisfy $n-2m\geq1$
so that trimming does not remove all $n$ data points. We first develop
an estimator for a fixed $\lambda$ and defer discussions on the choice of
$\lambda$ to Section \ref{sec:choice-trim-rate}.

By Proposition \ref{thm:symmetry}, $\overline{\epsilon}_{n\lambda}(\theta^{*})$
has an expected value of 0, so we can estimate $\theta^{*}$ by solving
\begin{align}
\overline{\epsilon}_{n\lambda}(\theta) & =0. \label{eq:trimmed-match-eq}
\end{align}
When multiple roots exist, we choose the one that minimizes
\begin{align}
D_{n\lambda}(\theta) & \equiv \frac{1}{n-2m}\sum_{i=m+1}^{n-m}|\epsilon_{(i)}(\theta)+\epsilon_{(n-i+1)}(\theta)|,\label{eq:symmetry-deviation-from-0}
\end{align}
a statistic which measures the symmetric deviation from 0
\citep{dhar2012derivatives}. More formally, we can express
this estimator as:
\begin{align}
\hat{\theta}_{\lambda}^{(trim)} & =\arg\min_{\theta}\{D_{n\lambda}(\theta):\overline{\epsilon}_{n\lambda}(\theta)=0\}.\label{eq:trimmed-match}
\end{align}

If $\lambda=0$, then no trimming takes place and
$\hat{\theta}_{\lambda}^{(trim)}$ coincides with
the empirical estimator $\hat{\theta}^{(emp)}$ from \eqref{eq:empirical}.
Although no simple closed
form for $\hat{\theta}_{\lambda}^{(trim)}$ exists when $\lambda>0$,
it is straightforward to show that
\begin{align}
\hat{\theta}_{\lambda}^{(trim)} & =\frac{\sum_{i\in\mathcal{I}}Y_{i}}{\sum_{i\in\mathcal{I}}X_{i}},\label{eq:interpret-trim-match}
\end{align}
where $\mathcal{I}$ is the set of $n-2m$ untrimmed indices of
$\epsilon_{i}(\theta)$ used in the calculation of
$\overline{\epsilon}_{n\lambda}(\hat{\theta}_{\lambda}^{(trim)})$ and thus
$\mathcal{I}$ depends on $\hat{\theta}_{\lambda}^{(trim)}$. Note that if the two
geos in the $i$th pair are perfectly matched in terms of the uninfluenced response,
then $\epsilon_{i}(\theta^{*})=0$.
Therefore, $\hat{\theta}_{\lambda}^{(trim)}$ has a nice interpretation: it trims
the poorly matched pairs in terms of the $\epsilon_{i}(\theta^{*})$ values and
estimates $\theta^{*}$ using only the well-matched untrimmed pairs.
Consequently, in this paper, we refer to $\hat{\theta}_{\lambda}^{(trim)}$ as
the ``Trimmed Match'' estimator.

It is worth emphasizing that Trimmed Match directly estimates $\theta^{*}$ without
having to estimate either the incremental response or the incremental spend.
Moreover, the point estimate is calculated
after trimming the pairs that are poorly matched in terms of the
$\epsilon_{i}(\hat{\theta}_{\lambda}^{(trim)})$ values
rather than the pairs which are poorly matched with respect
to the differences in their response $Y_{i}$ or ad spend $X_{i}$.
Indeed, consider an alternative trimmed estimator which does not directly
estimate $\theta^{*}$, but instead first separately calculates a trimmed
mean estimate of the average incremental response and a trimmed mean estimate of
the average incremental ad spend, and then takes their ratio:
\[
\frac{\sum_{i\in\mathcal{I}_{Y}}Y_{i}/|\mathcal{I}_{Y}|}{\sum_{i\in\mathcal{I}_{X}}X_{i}/|\mathcal{I}_{X}|},
\]
where the sets $\mathcal{I}_{Y}$ and $\mathcal{I}_{X}$ denote
the indices of the untrimmed pairs used for estimating the incremental response
and the incremental ad spend, respectively, and where the two sets will generally
not be identical.  Note, however, that
this is not a desirable estimator for $\theta^{*}$ since its numerator and
denominator may not even yield an unbiased estimate of either the average incremental response
or the average incremental spend, respectively, as neither $\{Y_{i}:1\leq i\leq n\}$ nor
$\{X_{i}:1\leq i\leq n\}$ is expected to follow a symmetric distribution
even if all of the geo pairs are perfectly matched in terms of
their ``uninfluenced responses''.

Finally, it is also interesting to note the connection between trimming poorly
matched pairs and the theory presented in \citet{small:rosenbaum:2008} which
shows that a smaller study with a stronger instrument is likely to be more
powerful and less sensitive to biases than a larger study with a weaker
instrument.  These arguments were later supported empirically by
\citet{baiocchi2010building}, who studied a similar effect ratio in the form of
\eqref{eq:theta}
and showed that optimally removing about half of the data in order to
define fewer pairs with similar pre-treatment covariates but with
stronger instrument
resulted in shorter confidence intervals and more reliable conclusions.
Our Trimmed Match method also identifies and trims poorly matched
pairs, but does not rely on pre-treatment covariates.

\subsection{Confidence Interval}
\label{subsec:trimmed-match-CI}

Define the studentized trimmed mean statistic
\citep{tukey1963less} with respect to $\{\epsilon_{i}(\theta):1\leq i\leq n\}$
as follows:
\begin{equation}
T_{n\lambda}(\theta)=\frac{\bar{\epsilon}_{n\lambda}(\theta)}{\hat{\sigma}_{n\lambda}(\theta)/\sqrt{n-2m-1}},\label{eq:studentized_trimmed_mean}
\end{equation}
where
\[
\begin{split}\hat{\sigma}_{n\lambda}^{2}(\theta) & =\frac{m\left[\epsilon_{(m+1)}(\theta)\right]^{2}+\sum_{i=m+1}^{n-m}\left[\epsilon_{(i)}(\theta)\right]^{2}+m\left[\epsilon_{(n-m)}(\theta)\right]^{2}-n\left[\overline{w}_{n\lambda}(\theta)\right]^{2}}{n-2m}\end{split}
\]
is the winsorized variance estimate for $\bar{\epsilon}_{n\lambda}(\theta)$,
and
\[
\bar{w}_{n\lambda}(\theta)=\frac{m\cdot\epsilon_{(m+1)}(\theta)+\sum_{i=m+1}^{n-m}\epsilon_{(i)}(\theta)+m\cdot\epsilon_{(n-m)}(\theta)}{n}
\]
is the winsorized mean of $\epsilon_{i}(\theta)$'s. The Trimmed Match
confidence interval is constructed by determining the minimal interval containing
all $\theta\in\mathcal{R}$  satisfying
\begin{align}
|T_{n\lambda}(\theta)| & \leq c\label{eq:T-confidence-interval},
\end{align}
where the threshold $c$ is chosen such that $\mathbb{P}\left(\left|T_{n\lambda}(\theta^{*})\right|\leq c\right)=1-\alpha.$

Under mild conditions,  $T_{n\lambda}(\theta^{*})$
approximately follows a Student's $t$-distribution with $n-2m-1$ degrees of
freedom, and
we therefore set $c$ to be the $(1-\alpha/2)$ quantile of this distribution.
Alternatively, one can also choose the threshold $c$ by using Fisher's
randomization test approach (see, for example, \citet{rosenbaum2002covariance}
and \citet{ding:etal:2016}) and relying on the fact that the distribution of
$\epsilon_{i}(\theta^{*})$ is symmetric about zero for each $i$.

However, when constructing the confidence interval, it is also important to
recognize that the
trim rate $\lambda$ is unknown in practice. Later, in Section \ref{sec:choice-trim-rate},
we propose a data-driven estimate of this trim rate which can
be used by plug-in to construct the confidence interval, although
such an interval may suffer from undercoverage in finite
samples as it ignores the uncertainty associated with estimating this tuning
parameter \citep{ding:etal:2016}. Interestingly, however, our numerical studies
in Section \ref{sec:Simulation-and-sensitivity}
suggest that the empirical coverage of the confidence intervals constructed using the
estimated trim rates are often quite close to the
nominal level even when $n$ is small---a finding which is consistent
with the observation that the studentized trimmed mean belongs to
the class of ``less vulnerable confidence and significance procedures''
for the classical location problem \citep{tukey1963less}.

\section{Data-driven Choice of the Trim Rate $\lambda$\label{sec:choice-trim-rate}}


For the location problem, \citet{jaeckel1971some} proposed minimizing the
empirical estimate of the asymptotic variance when choosing the trimmed mean's
trim rate $\lambda$, while \citet{hall1981large} proved
the general consistency of this approach. Similarly, we could choose the
trim rate $\lambda$ for Trimmed Match by minimizing an estimate of the
asymptotic variance of $\hat{\theta}_{\lambda}^{(trim)}$, which can be
derived by assuming independent sample for $(X, Y)$.\footnote{Some asymptotic analysis is reported in an earlier version at \url{https://arxiv.org/abs/1908.02922v2}.}
This may apply to the
scenario with no budget constraints, but not with budget constraints.

We note that the essential idea of \citet{jaeckel1971some} is to choose the trim
rate by minimizing the uncertainty of the point estimate measured by approximate
variance. To handle both scenarios without budget constraints and with budget
constraints, we extend this idea and propose to choose the trim rate by minimizing
the uncertainty of the point estimate measured by the width of the
$100(1 - \alpha_0)\%$ two-sided confidence interval previously defined in
Section \ref{subsec:trimmed-match-CI}, where $\alpha_0 \in (0, 1)$ is pre-specified.
Although different levels of $\alpha_0$ can be used,
our numerical studies in Section \ref{sec:Simulation-and-sensitivity} suggest
$\alpha_0 = 0.5$ to be a reliable choice in terms of its mean squared error
performance for both light and heavy tailed distributions. Hereafter we use $\hat{\lambda}$
to denote this data-driven trim rate using $\alpha_0=0.5$.


\section{Simulation and Sensitivity Analysis\label{sec:Simulation-and-sensitivity}}

In this section, we present several numerical simulations which evaluate the
performance and sensitivity of the Trimmed Match estimator
$\hat{\theta}^{(trim)}_{\lambda}$ defined by \eqref{eq:trimmed-match}.
We consider interferences between experimental units due to the presence of
budget constraint on the total incremental ad spend. To meet the requirement of
Assumption \ref{assm:SUTVA}, we consider the scenarios where the potential
ad spend under the control condition is pre-specified for each geo and will not
be affected by any geo assignment, but the potential ad spend under the
treatment condition may vary subjective to geo assignment, as
discussed in Section \ref{sec:A-model-free-framework}.
In particular, for simulations where
Assumption \ref{assm:common_iroas} holds, we investigate how
the choice of the trim rate $\lambda$ affects the performance of
$\hat{\theta}^{(trim)}_{\lambda}$ and, more broadly, we compare its
performance against the empirical estimator $\hat{\theta}^{(emp)}$ given
by \eqref{eq:empirical}, as well as the binomial sign-test estimator
$\hat{\theta}^{(binom)}$ and the Wilcoxon signed-rank-test estimator
$\hat{\theta}^{(rank)}$ defined in Section \ref{sec:binomial-test}.  Meanwhile, for
simulations where Assumption
\ref{assm:common_iroas} is violated, we investigate how
the level of deviation from Assumption \ref{assm:common_iroas}
affects the performances of these estimators.

For each simulation scenario, we first simulate the size of each geo $g=1,2,\ldots,2n$ as
\begin{align*}
z_{g} & =F^{-1}\left(\frac{g}{2n+1}\right),
\end{align*}
where $F$ controls the amount of geo heterogeneity in the population and is
taken to be either
a half-normal distribution, a log-normal distribution,
or a half-Cauchy distribution.  The geos are then paired based on these sizes (the largest two geos
form a pair, the third and fourth largest geos form a pair, and so on), and
afterwards the geos are randomized within each pair, which determines
whether a geo's control or treatment potential outcome is observed.

We list the detailed simulation steps as follows.
\begin{enumerate}[left=0pt]
\item[Step 1.] Potential ad spend and response under the control condition
according to a nonlinear relationship which is not affected by geo assignment: for $g\in \{1, \cdots, 2n\}$,
\[
  S_{g}^{(C)}=0.01 \times z_{g} \times (1 + 0.25 \times (-1)^g) \quad\text{and}\quad R_{g}^{(C)}=z_{g}.
\]
Let $B = 0.25 \times r \times \sum_{g=1}^{2n} S_g^{(C)}$ be the pre-specified budget for total incremental ad spend,
where $r > 0$ is a parameter controlling the intensity of incremental ad spend.

\item[Step 2.] Given a geo assignment denoted by $(\mathcal{T}, \mathcal{C})$
for the treatment and control groups respectively, the incremental ad spend for
each $g\in \mathcal{T}$ is proportional to their potential control spend:
\[
  \Delta_g^S = S_g^{(C)} \times B / \sum_{g' \in \mathcal{T}} S_{g'}^{(C)}.
\]

\item[Step 3.] Observed ad spend and response: For each $g\in \mathcal{C}$,
$S_g = S_g^{(C)}$ and $R_g = R_g^{(C)}$; For each $g\in \mathcal{T}$,
\[
  S_{g}=S_{g}^{(C)} + \Delta_g^S \quad\text{and}\quad R_{g}=R_{g}^{(C)}+\theta_{g}\times \Delta_g^S,
\]
where $\theta_{g}=\theta_0\times (1+\delta\times(-1)^{g})$ is the iROAS for
geo $g \in \{1, 2, \cdots, 2n\}$
with $\delta \in [0, 1]$ controlling the level of deviation from Assumption
\ref{assm:common_iroas}.
\end{enumerate}

The overall iROAS $\theta^*$ as defined in \eqref{eq:theta} can be rewritten as
$$\theta^* = \frac{\sum_{g=1}^{2n} \theta_g \times \Delta_g^S}{\sum_{g=1}^{2n} \Delta_g^S}$$
which may not be well-defined any more when potential outcomes depend on geo assignment.
When Assumption \ref{assm:common_iroas} holds,
i.e. $\theta_g \equiv \theta_0$ $\forall g\in{1,\cdots,2n}$,
then $\theta^* \equiv \theta_0$ for any assignment.
When Assumption \ref{assm:common_iroas} is violated, to get around the ill-definition,
we may assume a virtual experiment where all $2n$ geos were assigned to treatment with
a doubled total incremental budget, i.e. $2B$, and where the incremental budget for each
geo is still proportional to their potential control spends. It is easy to show that for this
virtual experiment
$\Delta_g^S = 0.5 \times r \times S_g^{(C)}$ for each $g$, and that the overall iROAS
can be simplified to a well-defined static quantity:
$$\theta^* = \theta_0 + \delta \times \theta_0 \times \frac{\sum_g z_g \cdot (0.25 + (-1)^g)}{\sum_g z_g \cdot (1 + 0.25 \times (-1)^g)}$$
which will be treated as the source of truth.

To summarize, the simulation parameters which are allowed to vary from scenario to scenario are the
number of geo pairs $n$, the distribution $F$ controlling the amount of geo heterogeneity, the
iROAS scale $\theta_0$, the intensity of the incremental ad spend $r$, and the level of deviation $\delta$
from Assumption \ref{assm:common_iroas}.  Within each scenario, we then simulate $K=10,000$
random assignments---a process that
determines which bivariate outcome $(S_{g},R_{g})$ is actually observed
for each geo $g$, and also the observed differences $(X_{i},Y_{i})$ as defined in
\eqref{eq:paired-diff} for each geo pair $i$.  Note that this assignment mechanism is the only
source of randomness within each of our simulations.

We summarize the results for $n=50$ and $\theta_0=10$ for each scenario reported in this section,
although we note that other simulation parameters (e.g. $n=25$) yielded similar conclusions.
The performance of an estimator's point estimate $\hat{\theta}$
is evaluated in terms of its root mean square error
\begin{align*}
RMSE(\hat{\theta}) & =\sqrt{\frac{1}{K}\sum_{k=1}^{K}\left(\hat{\theta}^{(k)}-\theta^{*}\right)^{2}}
\end{align*}
and its bias
\begin{align*}
Bias(\hat{\theta}) & =\frac{1}{K}\sum_{k=1}^{K}\hat{\theta}^{(k)}-\theta^{*},
\end{align*}
where $\hat{\theta}^{(k)}$ is the estimated value of $\theta^{*}$
from the $k$th replicate.  Meanwhile, the performance of an estimator's $100(1-\alpha)\%$ confidence
interval $(\hat{\theta}_{\alpha / 2}, \hat{\theta}_{1 - \alpha / 2})$ is measured
in terms of its one-sided power
\begin{align*}
Power(\hat{\theta}) & =\frac{1}{K}\sum_{k=1}^{K}I(\hat{\theta}^{(k)}_{\alpha / 2} > 0)
\end{align*}
and its two-sided empirical coverage
\begin{align*}
Coverage(\hat{\theta}) & =\frac{1}{K}\sum_{k=1}^{K}I(\hat{\theta}^{(k)}_{\alpha / 2} < \theta^{*} < \hat{\theta}^{(k)}_{1 - \alpha / 2}),
\end{align*}
where $(\hat{\theta}^{(k)}_{\alpha / 2}, \hat{\theta}^{(k)}_{1 - \alpha / 2})$ denotes the
confidence interval from the $k$th replicate.


\subsection{Performance Comparison When Assumption \ref{assm:common_iroas} Holds}

\begin{table}[t]
\centering \caption{Performance comparison w.r.t. RMSE (bias) when Assumption \ref{assm:common_iroas} holds}
\label{table:comparison_rmse_fixed_budget}
\scalebox{0.9}{
\begin{tabular}{c|c|ccccc}
\hline 

distribution & r & $\hat{\theta}^{(emp)}$ & $\hat{\theta}^{(binom)}$ & $\hat{\theta}^{(rank)}$ & $\hat{\theta}_{0.10}^{(trim)}$ & $\hat{\theta}_{\hat{\lambda}}^{(trim)}$ \tabularnewline

\hline 

 & 0.5 & \textcolor{red}{2.54 (0.29)} & \textcolor{red}{688.47 (1.20)} & \textcolor{red}{27.82 (0.64)} & \textcolor{red}{40.40 (0.78)} & 1.09 (-0.07) \tabularnewline

Half-Normal & 1.0 & 0.35 (0.05) & \textcolor{red}{490.91 (19.51)} & \textcolor{red}{5.30 (0.31)} & 0.51 (0.12) & 0.38 (-0.00) \tabularnewline

 & 2.0 & 0.16 (0.01) & \textcolor{red}{1.59 (0.94)} & 0.24 (0.04) & 0.19 (0.03) & 0.19 (-0.01) \tabularnewline

\hline 

 & 0.5 & \textcolor{red}{20.09 (0.64)} & \textcolor{red}{224.89 (-26.62)} & \textcolor{red}{111.33 (-0.16)} & \textcolor{red}{114.92 (1.07)} & 1.96 (0.10) \tabularnewline

Log-Normal & 1.0 & 0.86 (0.11) & \textcolor{red}{219.46 (-2.93)} & \textcolor{red}{7.84 (0.35)} & 0.60 (0.14) & 0.54 (0.04) \tabularnewline

 & 2.0 & 0.41 (0.03) & \textcolor{red}{1.65 (0.99)} & 0.28 (0.05) & 0.23 (0.03) & 0.24 (0.00) \tabularnewline

\hline 

 & 0.5 & 10.26 (-1.92) & \textcolor{red}{2143.22 (-6.85)} & \textcolor{red}{446.95 (6.96)} & \textcolor{red}{222.86 (-0.73)} & 5.20 (1.02) \tabularnewline

Half-Cauchy & 1.0 & \textcolor{red}{4.49 (-0.37)} & \textcolor{red}{299.33 (9.33)} & \textcolor{red}{2.25 (0.63)} & 1.03 (0.27) & 1.27 (0.21) \tabularnewline

 & 2.0 & \textcolor{red}{2.20 (-0.08)} & \textcolor{red}{1.80 (1.06)} & 0.47 (0.12) & 0.34 (0.06) & 0.36 (0.02) \tabularnewline

\hline 

\end{tabular} 

}

\end{table}

\begin{table}[t]
\centering \caption{Comparison of power (empirical coverage) when Assumption \ref{assm:common_iroas} holds}
\label{table:comparison_power_fixed_budget}
\scalebox{0.9}{
\begin{tabular}{c|c|ccccc}
\hline 

distribution & r & $\hat{\theta}^{(emp)}$ & $\hat{\theta}^{(binom)}$ & $\hat{\theta}^{(rank)}$ & $\hat{\theta}_{0.10}^{(trim)}$ & $\hat{\theta}_{\hat{\lambda}}^{(trim)}$ \tabularnewline


\hline 

 & 0.5 & 86 (95) & 3 (95) & 52 (95) & 55 (95) & 88 (92) \tabularnewline

Half-Normal & 1.0 & 100 (89) & 9 (97) & 91 (93) & 99 (90) & 100 (87) \tabularnewline

 & 2.0 & 100 (89) & 100 (93) & 100 (90) & 100 (89) & 100 (86) \tabularnewline

\hline 

 & 0.5 & 16 (96) & 1 (94) & 50 (95) & 52 (95) & 60 (92) \tabularnewline

Log-Normal & 1.0 & 47 (93) & 9 (96) & 97 (92) & 100 (90) & 99 (88) \tabularnewline

 & 2.0 & 100 (93) & 100 (93) & 100 (90) & 100 (90) & 100 (87) \tabularnewline

\hline 

 & 0.5 & 0 (100) & 0 (96) & 7 (95) & 9 (96) & 13 (92) \tabularnewline

Half-Cauchy & 1.0 & 0 (100) & 28 (97) & 98 (91) & 98 (93) & 94 (92) \tabularnewline

 & 2.0 & 0 (100) & 100 (93) & 100 (90) & 100 (93) & 100 (89) \tabularnewline

\hline 

\end{tabular} 

} 

\end{table}

We first fix $\delta=0$ (i.e. Assumption \ref{assm:common_iroas} holds) to investigate the performance of the estimators as we vary
the geo heterogeneity
$$F \in \left\{\text{Half-Normal, Log-Normal, Half-Cauchy}\right\}$$ and
the incremental ad spend intensity $r \in \{0.5, 1, 2\}$.

Table 
\ref{table:comparison_rmse_fixed_budget}
summarizes the simulation results in
terms of each estimator's RMSE and bias.  Here we see that RMSE and bias of
every estimator improves as the intensity of the incremental ad spend $r$
increases, and we note that the rank test based estimator
$\hat{\theta}^{(rank)}$ is generally more efficient than the sign test based estimator
$\hat{\theta}^{(binom)}$, but both of them perform poorly
relative to the Trimmed Match estimator $\hat{\theta}_{\hat{\lambda}}^{(trim)}$.
In addition, recall that the Trimmed Match
estimator $\hat{\theta}^{(trim)}_{\lambda}$ coincides with the empirical
estimator $\hat{\theta}^{(emp)}$ when the trim rate $\lambda = 0$.  Thus, if we
focus specifically on the performance of the Trimmed Match estimator, we see
that some level of trimming can be beneficial when the geo sizes
are generated from the more heterogeneous Log-Normal and Half-Cauchy
distributions. It is interesting to note that the data-driven choice $\hat{\lambda}$
generally perform better than a fixed choice of the trim rate
$\lambda \in \{0, 0.10\}$,
even for the Half-Normal scenario at $r=0.5$.
For each scenario, RMSEs greater than 2 times the RMSE for the best
performed estimator are colored in red.
Obviously, among all the estimators, $\hat{\theta}_{\hat{\lambda}}^{(trim)}$
is the most robust for all the three distributions.

Meanwhile, Table \ref{table:comparison_power_fixed_budget}
summarizes the power and empirical
coverage for each estimator's accompanying $90\%$ confidence interval.
Indeed, although the table suggests that the Trimmed Match estimator
with the data-driven estimate $\hat{\lambda}$ of the trim rate can
suffer slightly from undercoverage---a result which agrees with our discussions
in Section \ref{subsec:trimmed-match-CI}---we note that this estimator also provides
considerably more power than $\hat{\theta}^{(emp)}$ and $\hat{\theta}^{(binom)}$
when there is a low ($r=0.5$) or moderate ($r=1.0$) level of incremental
ad spend, more power than $\hat{\theta}^{(rank)}$ when there is a low ($r=0.5$)
level of incremental ad spend, and is quite competitive for other scenarios.

\subsection{Sensitivity Analysis When Assumption \ref{assm:common_iroas} is Violated}

\begin{figure}[t]
 \centering
 \includegraphics[width=5in]{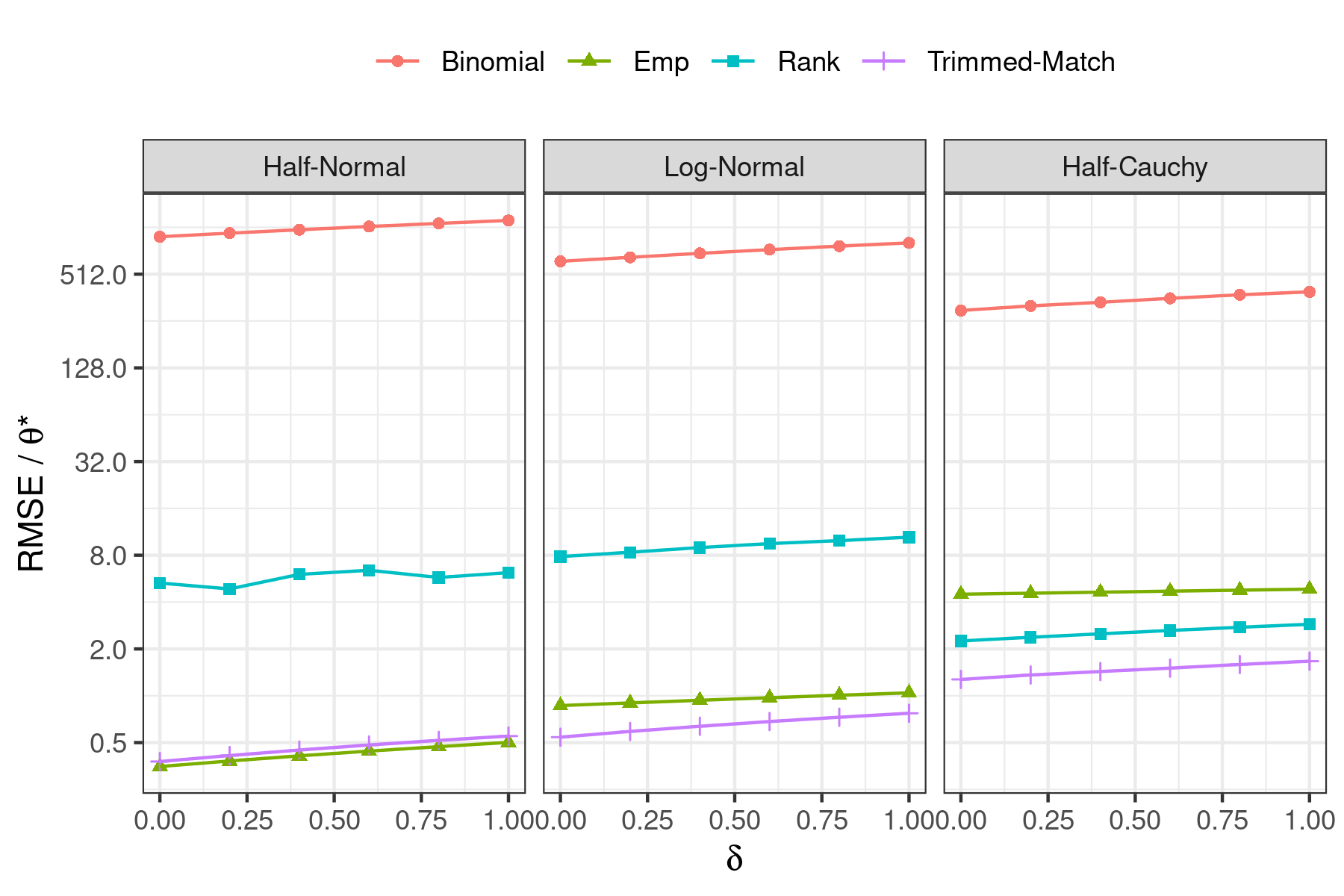}
 \caption{\label{fig:sensitivity_fixed_budget}Comparison of each estimator's performance
in terms of a scaled RMSE, where the $x$-axis $(\delta)$ quantifies the level of deviation
from Assumption \ref{assm:common_iroas}. Here the budget is fixed.}
\end{figure}

We now fix $r=1.0$ (a moderate level of incremental ad spend) and
evaluate the performance of the estimators when Assumption
\ref{assm:common_iroas} is violated---that is, the
geo-level iROAS are no longer the same. Instead, in these simulations,
half of the geos have an iROAS of $\theta_0(1-\delta)$ while the
other half have an iROAS of $\theta_0(1+\delta)$ according to Step 3.

Figure \ref{fig:sensitivity_fixed_budget} 
compares the performance of the estimators $\hat{\theta}^{(emp)}$,
$\hat{\theta}^{(binom)}$, $\hat{\theta}^{(rank)}$ and
$\hat{\theta}_{\hat{\lambda}}^{(trim)}$ in terms of a scaled RMSE.
Here we see that
$\hat{\theta}^{(trim)}_{\hat{\lambda}}$ significantly outperforms
$\hat{\theta}^{(binom)}$ and $\hat{\theta}^{(rank)}$ for all scenarios.
The empirical estimator $\hat{\theta}^{(emp)}$ slightly outperforms the
Trimmed Match estimator in the case of a half-normal
distribution, however, the Trimmed Match estimator significantly outperforms the
empirical estimator in the cases of the heavier tailed log-normal and
half-Cauchy distributions where there are stronger geo heterogeneity. Moreover,
we see that the Trimmed Match estimator still provides a useful estimate of
$\theta^{*}$ even when Assumption \ref{assm:common_iroas} is heavily violated
at $\delta=1$.

\section{Real Case Studies}
\label{sec:case-studies}

Next, we report real data analysis from three different geo experiments,
referred to as ``A'', ``B'' and ``C'', respectively, which were
run either in the United States or in Canada.
Each of these three experiments focused on a different business vertical, but they were
all designed by randomized matched geo pairs.
The number of geo pairs ranges from 60 to 105. Each experiment took a few weeks which
were split into two time periods: a test period during which the experiment
took place, and a cooldown period where the treatment geos were returned to
the control condition to account for potential lagged effects.  For each geo $g$,
$S_g$ and $R_g$ are the aggregated ad spend and aggregated
response over both time periods.

In experiments A and B, the advertisers wanted to measure the iROAS for
an improved ad format for their business where
geos assigned to the control group would run the business as usual using their
existing ad format, while geos assigned to the treatment group would adopt the
improved ad format. For either the control group or the treatment group,
the actual ad spend was much
lower than the budget pre-specified by the advertisers, and thus we expect no
interference, which implies Assumption \ref{assm:SUTVA}.

Experiment C was run to measure the iROAS of a new ad.
The ad would only be shown to geos in the treatment group, but not the control
group. The advertiser pre-specified a budget for the total ad spend.
After the experiment, the total ad spend for the control group was 0,
while for the treatment group was equal to
the budget, which implies strong interferences.
The potential control ad spend is 0 for each geo, and
thus Assumption \ref{assm:SUTVA} holds as discussed in Section
\ref{sec:A-model-free-framework}.

\begin{figure}[t]
 \centering
 \includegraphics[width=5.5in]{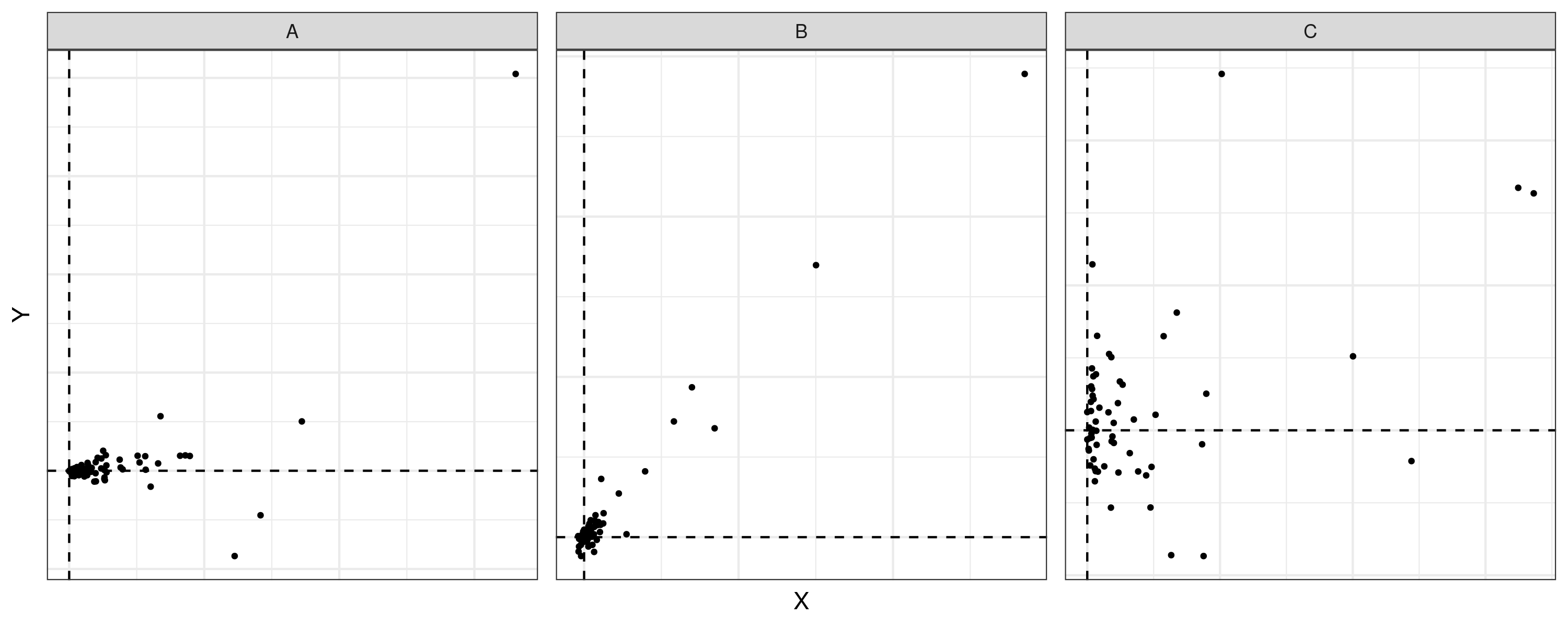}
 \caption{\label{fig:scatterplot}The scatter plot of $(X, Y)$ for each of the three real case studies, where
 the horizontal and vertical lines pass the origin but the detailed scales of both $X$ and $Y$ are removed to anonymize the experiments.}
\end{figure}

To illustrate the data heavy-tailedness, the scatter plots of $(X, Y)$ for
the three case studies are provided in Figure \ref{fig:scatterplot},
where the scales of both $X$ and $Y$ are removed in order to anonymize the experiments.
In Table \ref{tab:data-tailedness},
we report the kurtosis as a measure of heavy-tailedness for the empirical distributions of
\[
\{X_{i}:1\leq i\leq n\},\;\;\{Y_{i}:1\leq i\leq n\},\;\;\text{and}\;\;\{Y_{i}-\hat{\theta}_{\hat{\lambda}}^{(trim)}X_{i}:1\leq i\leq n\}.
\]
all of which are much larger than $3$, which is the kurtosis of any univariate normal distribution.

\begin{table}[t]
\centering \caption{\label{tab:data-tailedness} Summary of the three real case studies in terms of
the kurtosis (rounded to the nearest 10) of the empirical distributions,
the point estimates and confidence intervals obtained from different estimators
(rescaled by the point estimate $\hat{\theta}_{\hat{\lambda}}^{(trim)}$ to anonymize the experiments),
and the Trimmed Match's data-driven estimate $\hat{\lambda}$ of the trim rate.}

\resizebox{\textwidth}{!}{
\begin{tabular}{c||c|c|c|c|c|c|c|c}
\hline
Case  & Kurt($X$)  & Kurt($Y$)  & Kurt($\hat{\epsilon}$) & $\hat{\theta}^{(emp)}$ & $\hat{\theta}^{(binom)}$ & $\hat{\theta}^{(rank)}$  & $\hat{\theta}_{\hat{\lambda}}^{(trim)}$  & $\hat{\lambda}$ \tabularnewline
\hline \hline

A & 30 & 80 & 80 & 2.79          & 0.84          & 1.09         & 1.00        & 0.22 \\
  &    &    &    & [-1.26, 5.69] & [0.23, 1.81]  &[0.31, 1.85]  &[0.25, 1.74] &       \\
  \hline
B & 50 & 40 & 10 & 1.00          & 0.81          & 0.87         & 1.00        & 0.00 \\
  &    &    &    & [0.81, 1.12]  &[0.01, 1.06]   &  [0.56, 1.10]&[0.81, 1.12] &      \\
  \hline
C & 10 & 10 & 10 & 1.17          & 0.52          & 0.87         & 1.00        & 0.02 \\
  &    &    &    &  [0.14, 2.18] &[-1.50, 2.69]  & [-0.32, 1.97]&[-0.22, 1.94]&      \\

\hline
\end{tabular}
}
\end{table}

\begin{figure}[t]
 \centering
 \includegraphics[width=5.5in]{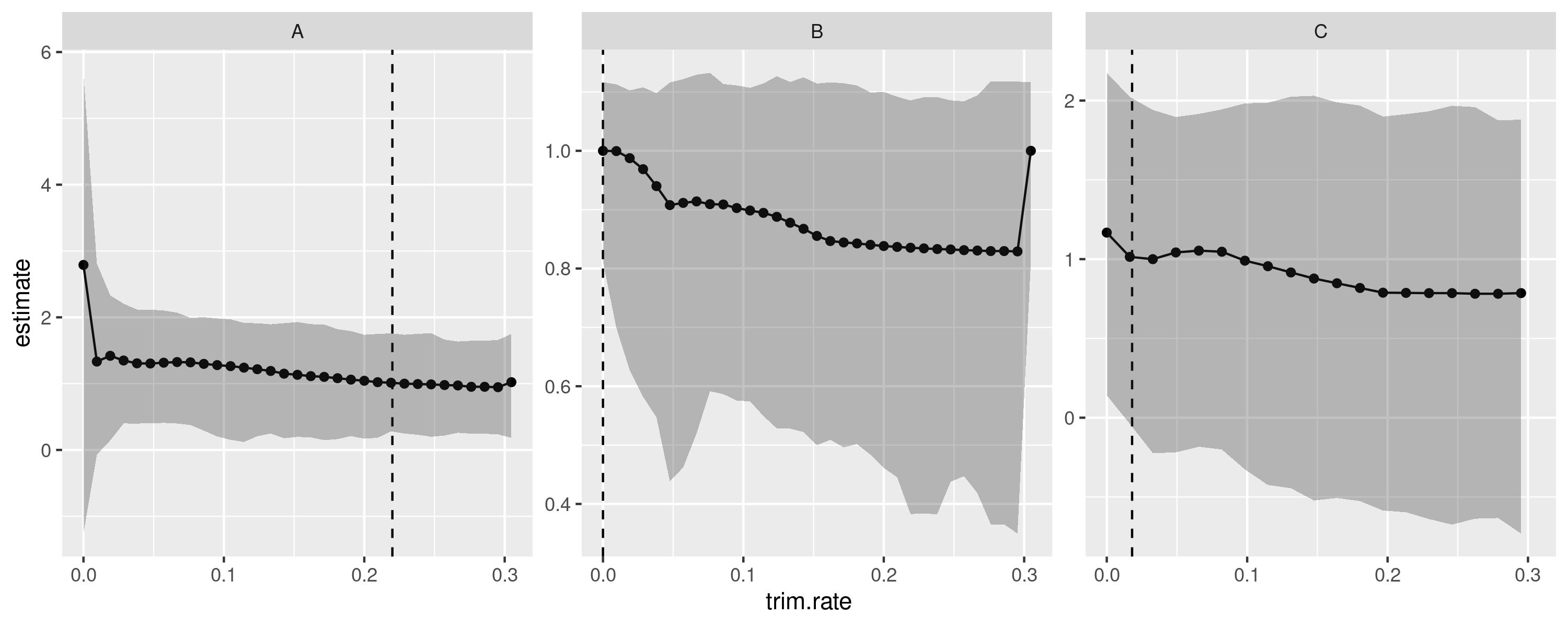}
 \caption{\label{fig:ci_bands}The Trimmed Match point estimates and confidence
intervals as a function of the trim rate $\lambda$ for each of the three real case
studies (rescaled by the point estimate $\hat{\theta}_{\hat{\lambda}}^{(trim)}$
to anonymize the experiments). The vertical dashed line corresponds to the
data-driven estimate $\hat{\lambda}$ of the trim rate.}
\end{figure}

Table \ref{tab:data-tailedness} also lists the point estimates and confidence
intervals for the empirical estimator $\hat{\theta}^{(emp)}$, sign test based estimator $\hat{\theta}^{(binom)}$,
rank test based estimator $\hat{\theta}^{(rank)}$, and
Trimmed Match estimator $\hat{\theta}^{(trim)}_{\hat{\lambda}}$.
Here we see that $\hat{\theta}^{(rank)}$ and
$\hat{\theta}^{(trim)}_{\hat{\lambda}}$ yield similar results
except in experiment B, where the confidence interval from the rank 
based estimator is almost 74\% wider than Trimmed Match.
Meanwhile, the sign test based estimator gives confidence intervals
that are even wider for all experiments by a considerable
amount.  It is also interesting to note that the data-driven choice
of Trimmed Match's trim rate results in no trimming (coinciding with
the empirical estimator $\hat{\theta}^{(emp)}$) for case B
despite the heavy-tailedness of the data, likely due to the
approximately linear relationship between $X$ and $Y$
(as shown in Figure \ref{fig:scatterplot})
so that trimming a geo pair with a large $|\epsilon_{i}(\theta^{*})|$ may not
necessarily reduce the variance.  As experiment A demonstrates, however,
the empirical estimator can sometimes be very unreliable as it is heavily
affected by outliers (as shown in Figures \ref{fig:scatterplot} and
\ref{fig:ci_bands}).

Meanwhile, Figure \ref{fig:ci_bands} plots the Trimmed Match point estimate and
confidence interval as a function of the trim rate $\lambda$. Here in case A,
we note the significant reduction in the size of the confidence intervals
relative to the empirical estimator $\hat{\theta}^{(emp)}$, i.e. no trimming.

\begin{figure}[ht]
 \centering
 \includegraphics[width=5.5in]{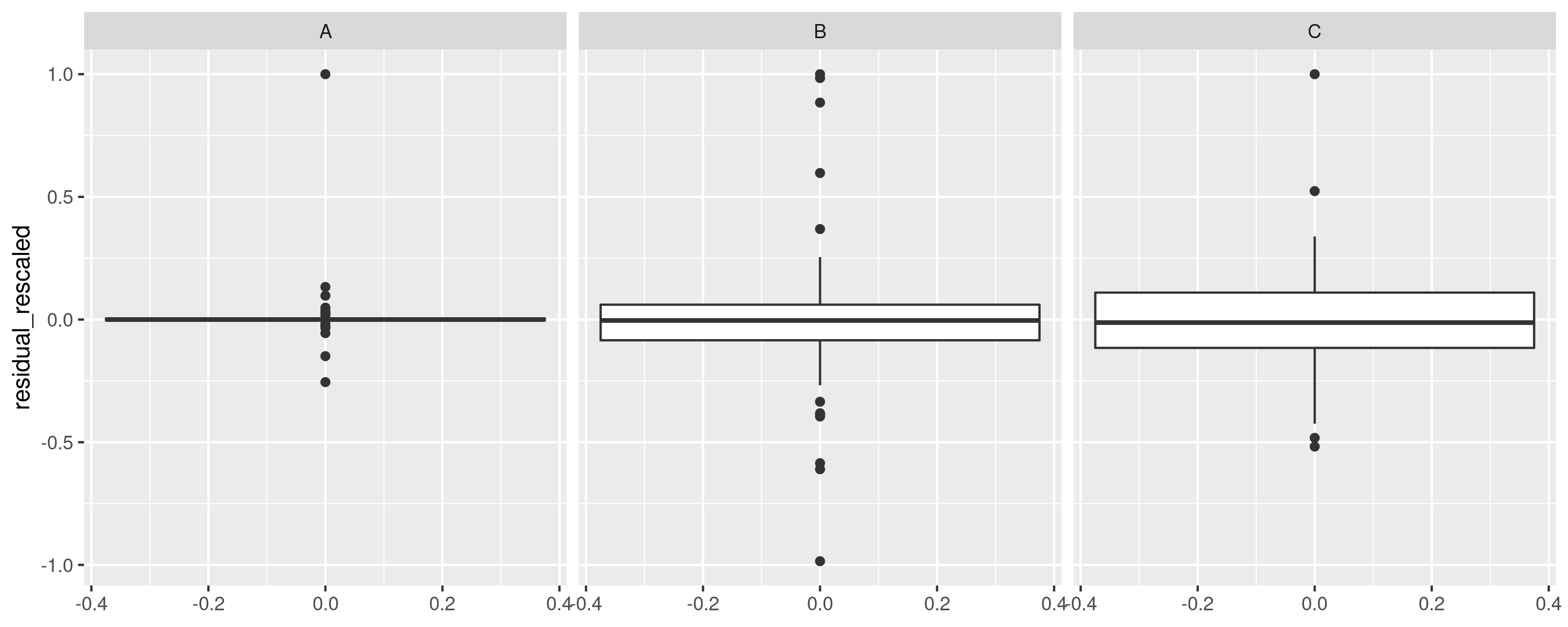}
  \caption{\label{fig:residual_boxplot}Boxplots of the residuals, where the residuals are rescaled by their maximum absolute value for each case separately.}
\end{figure}

We also investigate whether the real data are incompatible with the statistical
framework that we developed in Section \ref{sec:A-model-free-framework} under
Assumption \ref{assm:common_iroas}, which assumes that the geo-level iROAS
$\theta_{g}$ are all equal to one another.  Recall from
Proposition \ref{thm:symmetry} that the distribution
of the residuals $\left\{\epsilon_{i}(\theta^{*}) : 1 \leq i \leq n \right\}$
is symmetric about 0, where $n$ is the number of geo pairs.  Therefore, we expect
$\{\epsilon_{i}(\hat{\theta}_{\hat{\lambda}}^{(trim)}):1\leq i\leq n\}$
to be approximately symmetric about 0 as well---a null hypothesis which we can
test by using the Wilcoxon signed-rank test. The corresponding p-values\footnote{Note that
the p-values are based on the estimated $\theta^*$ (instead of the true values
which are unknown) and thus may not be accurate.}
are
0.66, 0.55 and 0.85 for the above three real case studies, which suggest that
the real data are not incompatible with Assumption \ref{assm:common_iroas}.
Figure \ref{fig:residual_boxplot} shows the boxplots of the fitted residuals for
the three cases respectively, which illustrate the heavy-tailedness as well as
approximate symmetry.

\section{Discussion\label{sec:Discussion}}

In this paper, we introduced the iROAS estimation problem in online advertising
and formulated a novel statistical framework for its causal inference
in a randomized paired geo experiment design which is often complicated by
the issues of small sample sizes, geo heterogeneity, and
interference due to budgetary constraints on ad spend. Moreover,
we proposed and developed a robust, distribution-free, and easily interpretable
Trimmed Match estimator which adaptively trims poorly matched geo pairs.
In addition, we devised a data-driven choice of the trim rate which extends
Jaeckel's idea but does not rely on asymptotic variance approximation, and
presented numerical studies showing that Trimmed Match is often more efficient
than alternative methods even when some of its assumptions are violated.

Several open research questions of considerable interest remain
such as 1) using Trimmed Match to improve the design of matched pairs experiments,
2) using covariates to further improve the estimation
precision \citep{rosenbaum2002covariance},
3) estimating the geo-level iROAS, and 4) further investigation of the choice
of the trim rate and corresponding asymptotic analysis,
e.g. using sample split \citep{klaassen1987consistent, nie2017quasi}.

Finally, although this paper focused on the estimation of the iROAS of online
advertising in geo experiments, we note that Trimmed Match can also be applied
to matched pairs experiments in other areas where the ratio of two causal estimands
is of interest (e.g., the incremental cost-effectiveness ratio \citep{chaudhary.stearns.1996,bang.zhao.2014};
see Chapter 5.3 of \citet{rosenbaum.book.2019} for more examples).

\begin{appendix}

\section{Fast Computation of Trimmed Match\label{sec:trimmed-match-comp}}

Recall from Section \ref{subsec:trimmed-match-point}, that obtaining the Trimmed
Match point estimate $\hat{\theta}_{\lambda}^{(trim)}$ requires finding all roots
of the trimmed mean equation
\eqref{eq:trimmed-match-eq}.  Moreover, recall that this computation is trivial
when $\lambda = 0$ as $\hat{\theta}_{\lambda}^{(trim)}$ just corresponds to the
empirical estimator given by \eqref{eq:random-empirical}.  Therefore, in the
remainder of this section, we focus on the computation with a fixed trim rate
$\lambda>0$.

Although \eqref{eq:interpret-trim-match} implies that calculating
$\hat{\theta}_{\lambda}^{(trim)}$ is straightforward once its
corresponding set of $n-2m$ untrimmed indices $\mathcal{I}$ is known,
$\mathcal{I}$ is generally \textit{a priori} unknown as
it depends on $\hat{\theta}_{\lambda}^{(trim)}$.  One could, at least in theory, check all
possible subsets of size $n-2m$, but this brute force approach requires the evaluation
of ${n \choose 2m}$ such subsets and would be computationally too expensive to be usable
in practice when $m$ is large.  However, by instead considering how the ordering
of the values in the set $\{\epsilon_{i}(\theta):1\leq i\leq n\}$ changes as
a function of $\theta \in \mathcal{R}$---in particular, by enumerating all
possible values of $\theta$ at which this ordering changes---we are able
to devise an efficient $O(n^{2}\log n)$ algorithm for
finding all of the roots of \eqref{eq:trimmed-match-eq}, which are required
by \eqref{eq:trimmed-match}.

Following \eqref{eq:paired-diff}, let $\{(x_{i},y_{i}):1\leq i\leq n\}$
be the differences in the ad spends and responses that are observed
from a randomized paired geo experiment.
For notational simplicity, assume
that $\{(x_{i},y_{i}):1\leq i\leq n\}$ is ordered such
that $x_{1}<x_{2}<\ldots<x_{n}$.

\begin{lem}
\label{lem:theta-range} For any two pairs of geos $i$ and $j$ such that $1\leq i<j\leq n$, let
\[
\theta_{ij}=\frac{y_{j}-y_{i}}{x_{j}-x_{i}}.
\]
Then
$\epsilon_{i}(\theta)<\epsilon_{j}(\theta)$ if and only if $\theta<\theta_{ij}$.
\end{lem}

Note that ties in $\{x_{i}: 1 \leq i \leq n\}$
rarely occur in practice;
when ties do occur, they can be broken by adding a small amount of random noise to
the $x_{i}$'s.  Lemma \ref{lem:theta-range}, whose proof is straightforward and
is omitted, allows us to efficiently solve the Trimmed Match equation defined
by \eqref{eq:trimmed-match-eq}.

\subsection{Solving the Trimmed Match Equation}

For ease of exposition, assume that $\{\theta_{ij}:1\leq i<j\leq n\}$ has been
ordered such that
$\theta_{i_{1}j_{1}}\leq\theta_{i_{2}j_{2}}\leq\ldots\leq\theta_{i_{N}j_{N}}$,
where $N=n(n-1)/2$.  Then, for any $k=1,2,\ldots,N-1$,
Lemma \ref{lem:theta-range} implies that the ordering of
$\{\epsilon_{i}(\theta):1\leq i\leq n\}$ is the same for all
$\theta\in(\theta_{i_{k}j_{k}},\theta_{i_{k+1}j_{k+1}})$ and, thus, the set of
untrimmed indices
\[
\mathcal{I}(\theta) \equiv\{1\leq i\leq n:\epsilon_{(m+1)}(\theta)\leq\epsilon_{i}(\theta)\leq\epsilon_{(n-m)}(\theta)\}
\]
must also be the same for all $\theta\in(\theta_{i_{k}j_{k}},\theta_{i_{k+1}j_{k+1}})$.
Moreover, Lemma \ref{lem:theta-range} also implies that
as $\theta$ increases and crosses a point $\theta_{i_{k}j_{k}}$, then
for any $1\leq i<j\leq n$, the ordering between $\epsilon_{i}(\theta)$ and
$\epsilon_{j}(\theta)$ changes if and only if $(i,j)=(i_{k},j_{k})$ or $(i,j)=(j_{k},i_{k})$.

Therefore, we can sequentially update the set of untrimmed indices $\mathcal{I}(\theta)$
based on what occurs as $\theta$ increases and crosses each point
$\theta_{i_{1}j_{1}}, \theta_{i_{2}j_{2}}, \ldots, \theta_{i_{N}j_{N}}$.
If $i_{k}, j_{k} \in \mathcal{I}(\theta)$ or
if $i_{k}, j_{k} \not\in \mathcal{I}(\theta)$, then $\mathcal{I}(\theta)$
remains unchanged;
if $i_{k} \in \mathcal{I}(\theta)$ but $j_{k} \not\in \mathcal{I}(\theta)$,
then we update
$\mathcal{I}(\theta)$ by replacing $i_{k}$ with $j_{k}$;
if $i_{k} \not\in \mathcal{I}(\theta)$ but $j_{k} \in \mathcal{I}(\theta)$,
then we update $\mathcal{I}(\theta)$ by replacing $j_{k}$ with $i_{k}$.
Pseudocode further describing this $O(n^{2}\log n)$ procedure is provided in
Algorithm \ref{alg:Solve-TrimmedMatch}.

\begin{center}
\begin{algorithm}
\caption{\label{alg:Solve-TrimmedMatch}Solving the Trimmed Match Equation \eqref{eq:trimmed-match-eq}}

Input: $\{(x_{i},y_{i}):1\leq i\leq n\}$ and trim rate $\lambda>0$; Let $m\equiv\nint{n\lambda}$.

Output: roots of \eqref{eq:trimmed-match-eq}.
\begin{enumerate}
\item Reorder the pairs $\{(x_{i},y_{i}):1\leq i\leq n\}$ such that $x_{1}<\ldots<x_{n}$;
Calculate $\{\theta_{ij}:1\leq i<j\leq n\}$ and order them such that
$\theta_{i_{1}j_{1}}<\theta_{i_{2}j_{2}}<\ldots<\theta_{i_{N}j_{N}}$.
(Break ties with negligible random perturbation if needed.)
\item Initialize the set of untrimmed indices
with
\begin{align*}
\mathcal{I} & = \{i:m < i \leq n-m\}
\end{align*}
Initialize $a = \sum_{i\in\mathcal{I}}y_{i}$,
$b = \sum_{i\in\mathcal{I}}x_{i}$, and
two ordered sets $\Theta_{1}=\{\}$ and $\Theta_{2}=\{\}$.
\item For $k=1,\ldots,N$:
\begin{enumerate}
\item If $i_{k}\in\mathcal{I}$ and $j_{k}\notin\mathcal{I}$, then update $\mathcal{I}, a, b$ as follows:
\begin{align*}
\mathcal{I} & \leftarrow\mathcal{I}+\{j_{k}\}-\{i_{k}\} \\
a & \leftarrow a+y_{j_{k}}-y_{i_{k}}\\
b & \leftarrow b+x_{j_{k}}-x_{i_{k}}
\end{align*}
and append $a/b$ to $\Theta_{1}$ and $\theta_{i_{k}j_{k}}$ to $\Theta_{2}$.
\item If  $j_{k}\in\mathcal{I}$ and $i_{k}\notin\mathcal{I}$, then update $\mathcal{I}$, $a$ and $b$ similar to (a), and append $a/b$ to $\Theta_1$ and $\theta_{i_{k}j_{k}}$ to $\Theta_2$.
\item Otherwise, continue.
\end{enumerate}
\item Append $\infty$ to $\Theta_{2}$, and output a subset of $\Theta_{1}$ as follows:
  For $k=1,\ldots,|\Theta_{1}|$, output $\Theta_{1}[k]$ iff $\Theta_{2}[k]\leq\Theta_{1}[k]\leq\Theta_{2}[k+1].$
\end{enumerate}
\end{algorithm}
\par\end{center}

\subsection{Computing the Confidence Interval}

Lemma \ref{lem:theta-range} also facilitates the calculation of the confidence
interval by reducing \eqref{eq:T-confidence-interval} to a quadratic inequality.
The specific details are omitted from this paper for conciseness but are available
from the authors upon request.

\subsection{Existence of $\hat{\theta}_{\lambda}^{(trim)}$}

From our discussions in this section, it is not necessarily obvious whether the
Trimmed Match point estimate $\hat{\theta}_{\lambda}^{(trim)}$ always exists.
However, the following theorem, whose proof is also omitted for conciseness, 
guarantees that it does indeed always exist as long as the trimmed mean of the
$x_i$'s is nonzero.

\begin{thm}
\label{thm:solution-exists} (Existence) Suppose that $\{(x_{i},y_{i}):1\leq i\leq n\}$
is ordered such that $x_{1}\leq x_{2} \leq \ldots\leq x_{n}$. Then:
\begin{enumerate}[1)]
 \item $\overline{\epsilon}_{n\lambda}(\theta)$ is a continuous function
with respect to $\theta\in\mathcal{R}$.
\item If $\sum_{i=m+1}^{n-m}x_{i}\neq 0$,
then $\overline{\epsilon}_{n\lambda}(\theta)=0$ has
at least one root.
\end{enumerate}
\end{thm}

\end{appendix}

\section*{Acknowledgements}
The authors would like to thank Art Owen and Jim Koehler for insightful early
discussion, Peter Bickel for the reference of Jaeckel's paper on the
choice of trim rate, Nicolas Remy, Penny Chu and Tony Fagan for the support,
Jouni Kerman, Yin-Hsiu Chen, Matthew Pearce, Fan Zhang, Jon Vaver, Susanna
Makela, Kevin Benac, Marco Longfils and Christoph Best for interesting
discussions, and the people who read and commented on the manuscript.
We appreciate Editor Beth Ann Griffin and
anonymous reviewers whose comments have helped improve the paper significantly.
All the figures are produced with the R package ggplot2 \citep{ggplot2}.

%
%
 


\bibliographystyle{imsart-nameyear} 
\bibliography{trimmed_match.bib}       


\end{document}